SURVEY

# A Survey of Algorithm Debt in Machine and Deep Learning Systems: Definition, Smells, and Future Work

EMMANUEL SIMON

CHIRATH HETTIARACHCHI

FATEMEH FARD

ALEX POTANIN

HANNA SUOMINEN

# A Survey of Algorithm Debt in Machine and Deep Learning Systems: Definition, Smells, and Future Work

EMMANUEL SIMON, School of Computing, Australian National University, Canberra, Australia
CHIRATH HETTIARACHCHI, School of Computing, Australian National University, Canberra, Australia
FATEMEH FARD, Computer Science Department, University of British Columbia, Kelowna, Canada
ALEX POTANIN, School of Computing, Australian National University, Canberra, Australia
HANNA SUOMINEN, School of Computing, Australian National University, Canberra, Australia and Department of Computing, University of Turku, Turku, Finland

**Abstract**. The adoption of Machine and Deep Learning (ML/DL) technologies introduces maintenance challenges, leading to Technical Debt (TD). Algorithm Debt (AD) is a TD type that impacts the performance and scalability of ML/DL systems. A review of 42 primary studies expanded AD's definition, uncovered its implicit presence, identified its smells, and highlighted future directions. These findings will guide an AD-focused study, enhancing the reliability of ML/DL systems.

## 1 Introduction

The rapid advancement of Machine Learning (ML) and Deep Learning (DL) technologies has been driven by progress in neural networks, data availability, and computing resources [55]. This has led to their widespread adoption across, e.g., healthcare, tourism, finance, and education sectors [31, 40, 89]. ML/DL systems (i.e., those that incorporate ML/DL algorithms or frameworks) in applications from an Artificial Intelligence (AI) chatbot called DeepSeek to voice assistants and self-driving cars are transforming how we interact with technology [13]. However, these complex systems inherit Technical Debt (TD) and the challenges of conventional software systems (i.e., those not incorporating any ML/DL), while introducing new ML-specific complexities that challenge standard software engineering practices [12, 73].

TD, first conceptualised by Cunningham [27], refers to expedient design choices that offer short-term benefits but accumulate long-term maintenance costs [4]. While TD maybe inevitable in software development, understanding its manifestations enables efficient mitigation strategies [59]. In ML/DL systems, TD takes on distinctive forms due to data dependencies, algorithmic complexity, and continuous model retraining requirements [73]. Among these, Algorithm Debt (AD) has emerged as a distinct TD type that impacts model performance [49]. AD in ML/DL systems is defined as a TD type that arises from inefficient implementation of algorithms, leading to limited scalability and model degradation.

Yet AD remains underexplored as a unique TD, often conflated with conceptually similar TD types or treated as an implicit concern rather than a focused research area. This conflation stems from insufficient conceptual clarity regarding what distinguishes AD from similar TD manifestations. While Liu et al. [49] first identified AD as a distinct category in DL frameworks, the boundaries between AD and Design, Model, Data, and traditional performance issues have not been articulated. This leads to conceptual ambiguity that makes practitioners and researchers unable to distinguish algorithmic inefficiencies from other optimisation concerns, undermining targeted mitigation efforts.

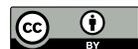

To clarify this boundary, understanding what makes AD a form of TD rather than merely a performance flaw is essential. AD represents a consciously incurred trade-off, where developers use suboptimal algorithmic implementations at the point of decision-making, with long-term implications. This decision distinguishes AD fundamentally, where developers sacrifice algorithmic efficiency (e.g., lower computational speed, higher memory usage, or reduced scalability) in exchange for immediate benefits such as faster development cycles, reduced implementation complexity, or meeting tight deadlines. The developers' awareness is often evidenced in code comments (e.g., TODO: optimise this algorithm), hyperparameter documentation, or commit messages that explicitly signal recognition of suboptimality. This deferral creates a debt structure, where the organisation accrues short-term gains such as faster time-to-market while incurring long-term maintenance liabilities such as increased computational costs, model degradation, and scalability constraints. The TD matures when the initial trade-off decisions require future remediation to meet production performance requirements or respond to system scaling demands.

This contrasts with related TD types such as Design, Model, Data, and performance concerns [5, 16]. Design Debt arises from structural trade-offs in system organisation such as insufficient modularity and tight coupling between components, rather than the algorithmic logic itself. Model Debt by contrast, concerns inappropriate, inflexible, or overfitted choices in ML model architecture, selection, or parameterisation, issues that arise from model-level decisions rather than algorithmic implementation. A model might be architecturally sound but trained with suboptimal algorithmic configurations. Data Debt encompasses issues of data quality, sufficiency, or representativeness, problems inherent to the data pipeline rather than algorithmic processing. Likewise, performance flaws describe inefficiencies identified through observation or testing, without prior developer awareness or explicit acknowledgment of the trade-off at implementation time. This distinction is important because it defines where and how mitigation efforts should be directed. Table 1, presents a summary of common TD types found in ML/DL systems according to Alves et al. [5]'s ontology and Liu et al. [49]'s characterisation.

Nevertheless, prior research has investigated TD in ML/DL systems without systematically isolating AD as a distinct phenomenon. Liu et al. [49] identified AD as one of seven TD types in DL frameworks, finding that it accounts for 5.6–20.7% of Self-Admitted TD (SATD) instances. Subsequent research by Bogner et al. [16] and Bhatia et al. [12] characterised TD manifestations in ML systems, yet few studies [76, 78], have focused exclusively on AD. The limited studies have left these questions unanswered: *How has AD been defined and what are the key components of a consolidated definition for AD in ML/DL systems? To what extent and in what ways has AD been addressed in ML/DL research? What are the main categories of AD smells in ML/DL systems?*

To address these gaps, the goal of this survey is to consolidate the literature on AD in ML/DL systems, with the aim to refine its definition, investigate its prevalence, identify its smells, and propose future directions. We used a hybrid search protocol [56, 57] to thematically analyse 42 papers retrieved from recommended software engineering libraries (including IEEE, ACM, Springer, and Science Direct), snowballing over Google Scholar [42]. We also used a dataset containing AD comments that were explicitly labelled from a previous study [49] to compliment our literature search. The findings from our survey include:

- *An expanded definition of AD in ML/DL systems.* We empirically refine AD's definition, grounded in literature and developer comments, uncovering a broader spectrum of impacts beyond performance, including model degradation and scalability. This provides a holistic perspective that aids communication among researchers and practitioners and promotes a shared understanding of AD.

A Survey of Algorithm Debt in Machine and Deep Learning Systems: Definition, Smells, and Future Work • 3

Table 1. Definition of different TD types with AD and Compatibility Debt highlighting an extension of Alves et al. [5]'s ontology by Liu et al. [49]. Model and Data Debt are adapted from Sculley et al. [73].

| **TD type** | **Definition** |
| --- | --- |
| Algorithm | A TD type that arises from the implementation of inefficient algorithm designs, leading to poor scalability and model degradation. |
| Architectural | A TD type that can arise from initial solutions that become sub-optimal as technologies and patterns evolve. |
| Code | Refers to the problems found in the source code which can affect negatively the legibility of the code making it more difficult to be maintained. |
| Compatibility | Refers to TD related to a project's immature dependencies on other projects, where the current implementation is a temporary workaround. |
| Defect | Refers to known defects, usually identified by testing activities or by the user. |
| Design | Indicates sub-optimal design, e.g., misplaced code, lack of abstraction, long methods, poor implementation, or workarounds on the usage of other internal functions. |
| Documentation | Refers to problems found in software project documentation and can be identified by looking for missing, inadequate, or incomplete documentation. |
| Implementation | Refers to trade-offs made concerning what requirements the development team needs to implement or how to implement them. |
| Model | Refers to issues arising from the use of inappropriate, inflexible model architectures, which can limit generalisation, robustness, or adaptability in ML/DL systems. |
| Data | Refers to problems associated with poor quality, insufficient, unrepresentative, or outdated data, which impact the effectiveness and reliability of ML/DL solutions. |

- *State of Research on AD in ML/DL Systems.* Our analysis reveals high implicit discussion of AD across primary studies, showing it is an overlooked phenomenon. This challenges the assumption that AD in ML/DL only emerged in 2020 and suggests historical misclassification with TD types such as Model and Data Debt, underscoring the need for its systematic recognition.
- *Taxonomy of AD smells.* We identify nine AD smells in ML/DL systems, linking them to TD manifestations but isolating them as AD-specific. This taxonomy frames AD as a combination of human and technical decisions, distinct from code-centric TD, and offers a foundation for future empirical and theoretical work.
- *A research agenda for future AD studies.* We propose a roadmap positioning AD as a dynamic, evolving concept. Future research should move from theoretical framing to practical strategies, including metrics, detection tools, and mitigation approaches, ensuring AD is addressed alongside other TD types.

## 2 Related Works

### 2.1 TD Evolution and Research Landscape

TD was first conceptualised by Cunningham [27] as suboptimal design choices that accrue over time, hindering maintainability and future evolution. Since this foundational work, TD research has evolved significantly across two complementary directions: foundational characterisation and systematic management. Foundational research established TD terminology and taxonomy. Alves et al. [5] and Avgeriou et al. [7] developed ontologies of TD and identified "smells" (namely, indicators of TD). Potdar and Shihab [69] introduced SATD by analysing source code comments in four Java projects, uncovering 62 textual patterns, though their scope excluded ML/DL systems.

Systematic reviews have consolidated TD knowledge across multiple dimensions. E.g., Lenarduzzi et al. [46] analysed TD prioritisation methods; Biazotto et al. [14] reviewed TD automation tools and challenges; and Ajibode et al. [1] analysed TD evolution forecasting approaches. These reviews confirmed that TD management remains an important challenge in traditional software systems, with architectural and design issues as the most prevalent causes.

Recent TD research has broadened to include diverse perspectives. Graf-Vlachy and Wagner [34] explored human factors such as developer personalities influencing TD accumulation across 37 projects. Verdecchia and Maggi [83] investigated TD evolution in microservices-based systems, identifying tight release dates, work pressure, and inexperience as common causes. Saeeda et al. [72] examined "non-TD" aspects such as good practices through a multivocal review, finding that addressing non-TD is more challenging than managing TD itself. Aldaeej and Alshayeb [3] replicated a global TD survey in Saudi Arabia with 48 practitioners, revealing that TD is not well-recognised in non-Western software industries suggesting geographic variations in TD awareness and management practices.

While extensive research addresses TD in traditional object-oriented software, ML/DL systems present distinct challenges. These distinct challenges are due to their data dependencies and complex algorithms.

## 2.2 TD in ML/DL Systems

Given the complex nature of ML/DL systems, they face unique TD challenges absent from traditional software. Sculley et al. [73] pioneered ML-specific TD research, identifying data dependencies and configuration issues as important risk factors. Building on this and research on TD by Alves et al. [5], Liu et al. [49] characterised TD across seven DL frameworks (TensorFlow, Keras, PyTorch, MXNet, Caffe, DL4J, CNTK), identifying seven TD types and notably introducing AD as a distinct category prevalent in ML/DL systems.

Subsequent research has refined understanding of ML/DL-specific TD through multiple complementary approaches. Bogner et al. [16] identified four additional ML-specific TD types (Data, Model, Configuration, Ethics Debt). Recent efforts have also further classified ML/DL-specific SATD, with Pepe et al. [67] categorising TD into different aspects of DL models some of which are technological factors (e.g., hardware or software libraries) and others related to procedural issues in DL like improper model usage, or suboptimal configuration settings. Ximenes [91] identified key factors contributing to TD in ML/DL, highlighting data processing and model training as major factors.

Furthermore, Bhatia et al. [12] manually analysed SATD across five ML domains, finding that data preprocessing and model generation logic are most susceptible to TD. Simon et al. [77] proposed automated SATD detection in DL frameworks using ML/DL techniques including logistic regression, Random Forest, RoBERTa, and Albert.

Beyond SATD, research on code quality and anti-patterns have been conducted in ML/DL systems. Cardozo et al. [20] analysed 24 Python ML repositories, identifying prevalent code smells such as long method and long chain that lead to reduced maintainability. Shivashankar and Martini [75] conducted a quantitative review of 56 papers on ML/DL engineering challenges, highlighting data engineering and model engineering workflows as the major challenges. Additionally, surveys on TD types in ML systems [70], have further refined TD classifications, linking TD to broader architectural and configuration challenges. While these studies collectively characterise TD manifestations in ML/DL systems, research on AD specifically its definition, prevalence, causes, and mitigation, remains limited to implicit discussions within broader TD studies.

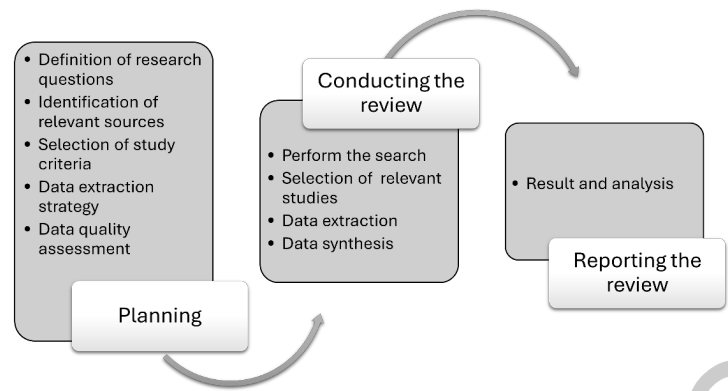

Fig. 1. Overall methodology for our survey process.

## 2.3 AD in ML/DL systems

AD emerged as a distinct TD type specifically in ML/DL systems, despite algorithmic inefficiencies existing in traditional software development. First introduced by Liu et al. [49] in their work on TD in DL frameworks, AD is defined as the suboptimal implementation of algorithm logic that negatively impacts system performance. While traditional software engineering addresses algorithmic inefficiencies under broader categories like performance optimisation or code quality, ML/DL applications present unique challenges including model selection trade-offs and hyperparameter optimisation. The presence of AD may have an impact on the outcomes produced by a ML/DL system, particularly in computationally intensive DL projects where suboptimal implementation of DL algorithms exacerbate performance issues. Given the heavy computational demands of DL frameworks, any inefficiency in algorithm logic can lead to resource wastage and degraded system performance.

Early research on AD also suggest the potential impact of AD for high-performance applications. E.g., Liu et al. [49] suggest that frameworks exhibiting higher levels of AD like Keras, might be less suitable for applications that are having intensive computational requirements. Overall, these findings highlight the need of identifying, managing, and mitigating AD to ensure that ML/DL systems can meet their intended performance and reliability goals.

## 2.4 Novelty

While TD research has undergone evolution and attention in recent years, particularly transitioning from a metaphorical concept to actionable engineering practices, no study has synthesised the role of AD (i.e., on system performance as per the definition by Liu et al. [49]) in ML/DL systems. Our survey addresses this gap, by consolidating the definition of AD and analysing prior studies. Beyond enriching the TD literature, our survey offers researchers insights into the AD smells, fostering discussions to enhance system reliability. These contributions serve as a foundational step for deeper investigations into AD, advancing TD research in the context of TD in ML/DL.

## 3 Methodology

We conducted a survey of literature using the hybrid-search review protocol, combining database search with snowballing [56, 57], to investigate AD in ML/DL systems. Before we selected this protocol, we considered the rapid review [21] as an alternative methodology. Rapid reviews are conducted to bring research evidence to practice within a short period, often at the cost of rigour and reproducibility. With our aim to consolidate the literature on AD in ML/DL systems to refine its definition, investigate its prevalence, identify its smells, and propose future directions, the rapid review approach was not suitable. The hybrid-search protocol that uses multiple reviewers and quality assessments was more suitable for addressing our objectives. Figure 1, illustrates the overall process of our survey, detailing the steps from the planning to reporting of the results.

### 3.1 Research Questions (RQs)

To achieve our goal, we defined three RQs as follows:

**RQ1: How has AD been defined and what are the key components of a consolidated definition for AD in ML/DL systems?** Emergent topics often have ambiguous definitions being proposed [28]. A clear and consolidated definition of AD, derived from a systematic survey of academic literature [43], is essential for establishing a shared understanding of this concept, enabling a focused research.

**RQ2: To what extent and in what ways has AD been addressed in ML/DL research?** Given that AD is underexplored in ML/DL systems, the aim of this RQ is to investigate how prior studies addressed AD-related issues in literature.

**RQ3: What are the main indicators of AD smells in ML/DL systems?** The aim of this RQ is to identify the main categories of AD smells (i.e., defined as literature-reported indicators of suboptimal algorithmic choices) in ML/DL systems. Identifying these smells reveals the practical manifestations of AD reported in the literature, building on the definition from RQ1, presenting AD as a distinct TD type, and provides motivation for proposing effective mitigation strategies.

### 3.2 Control Papers

To conduct this survey, we started by reading seven control papers (CP01–CP07) shown in Table 2, selected by a software engineering expert with over ten years of experience in TD research and industry experience in ML. The expert selected papers that are relevant to TD in ML/DL systems, where the papers either explicitly mentioned AD or related concepts (highlighting algorithmic inefficiencies). Papers pre-dating AD's first mention (2020) were included for their foundational insights into TD issues that were later associated with AD in ML/DL systems.

For example, CP01 explored hidden TD in ML systems, implying AD through data dependency and configuration challenges; CP02 analysed bugs in ML programs, hinting at AD-related inefficiencies. CP03 (Liu et al. [49]) investigated TD in DL frameworks and uncovered AD. CP04 investigated the emergence and management of TD in AI-based systems. CP05 discussed ML refactorings, and CP06 investigated code smells in ML systems. CP07 studied SATD in ML systems. A detailed description of each control paper is presented in our online repository[1].

### 3.3 Search and Selection

We followed the guidelines for conducting literature review to search and select relevant papers [42]. The specific steps based on these guidelines are illustrated in Figure 2.

*3.3.1 Search String and Query.* To construct the search string we used to retrieve relevant studies, we first identified our search domain by defining what should be considered as an ML/DL system. We adopted the definition by Bach et al. [9], recognising ML/DL systems as encompassing ML/DL algorithms and applications incorporating ML/DL components. This definition served as a foundation to identifying contexts where AD might arise in ML/DL systems.

---

[1]https://drive.google.com/drive/folders/1W4kcQBrVPZwrtG9vLcCnZURdxUOThNan

Table 2. Set of control papers (CP01–CP07) used for the study; the fourth column indicates if AD was directly mentioned in the paper or implied. Direct indicates explicit mention of AD, while Implied AD refers to discussions of related algorithmic inefficiencies or ML-specific TD issues without naming AD.

| Code | Author | Paper Title | AD Mention | Publication Year |
| --- | --- | --- | --- | --- |
| CP01 | Sculley et al. [73] | Hidden TD in ML systems | Implied | 2015 |
| CP02 | Sun et al. [79] | An Empirical study on real bugs for ML programs | Implied | 2017 |
| CP03 | Liu et al. [49] | Is using DL frameworks free? Characterising TD in DL frameworks | Direct | 2020 |
| CP04 | Bogner et al. [16] | Characterising TD and antipatterns in AI-based systems: A systematic mapping study | Implied | 2020 |
| CP05 | Tang et al. [80] | An Empirical study of refactorings and TD in ML systems | Implied | 2021 |
| CP06 | Van Oort et al. [81] | The prevalence of code smells in ML projects | Implied | 2021 |
| CP07 | OBrien et al. [64] | 23 shades of self-admitted TD: An empirical study on ML software | Implied | 2022 |

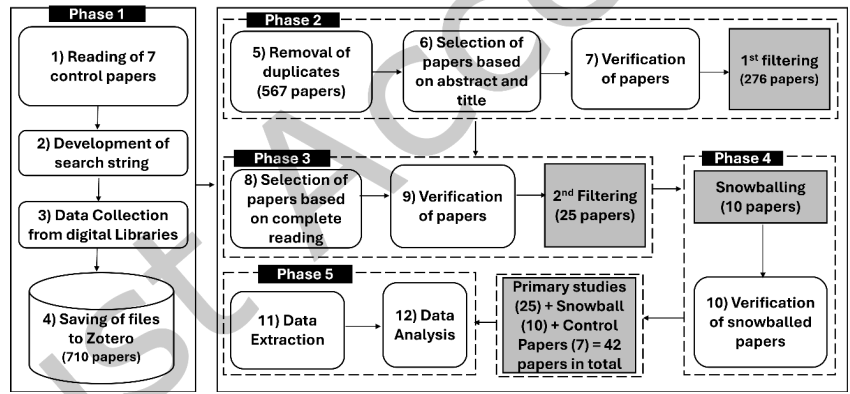

Fig. 2. Overview of the search and selection process detailed in Section 3.3

We crafted the search string to retrieve a wide range of relevant literature on TD and AD, by incorporating terms we identified from the initial control papers (CP01 - CP07). These terms included "code smells", "refactor", "research software", and "refactor*", all serving a distinct purpose. We included code smells since they are recognised as indicators of TD and could therefore indicate AD-related issues within ML/DL code [30, 66], particularly important to our investigation of the AD smells in RQ3. We added the terms "scientific software" and "research software" to capture ML/DL studies within academic and research contexts. The inclusion of "refactor" was important for identifying studies related to ML/DL software maintenance. We used the asterisk character (∗) to account for variations of the terms.

To capture the diverse terminology used in ML/DL domain, we included "artificial intelligence", "AI", "Deep Learning", "DL", "Machine Learning", and "ML". These terms are used interchangeably in research on ML/DL systems [16]. With our primary emphasis on AD, we explicitly included "Algorithm Debt" and "AD" in our search string, similar to the approach taken when searching for TD types like Architecture Debt [11]. An important component of our search string was the inclusion of "Hidden Technical Debt". This term specifically targets publications that reference the seminal work on TD in ML systems by Sculley et al. [73], which was foundational to the understanding of TD in ML systems (critical to our domain of ML/DL systems and RQ2).

To maximise the effectiveness of the search and avoid missing relevant literature, we used several strategies. We used the boolean operator "AND" so that the search can include all of the terms from both TD and ML/DL categories, and "OR" to search for any of the terms [19]. Since "AD" is not a common term in publication titles, we applied the search on both title, abstract, and body (full text).

Following systematic review guidelines [41], we used the control papers (see Section 3.2 and Table 2) to validate the effectiveness of our search strings. Taking inspiration from [14], we experimented with numerous variations of the search string on Google Scholar. This process involved optimising and refining the search query based on initial analyses and evaluations of the returned content. Specifically, our aim was to identify a set of search strings that would produce more than half of the initial control papers within the first few pages [14] from the search process.

At the end, the final query was:

> ("Technical Debt" OR "TD" OR "code debt" OR "code smells" OR refactor* OR "Algorithm Debt" OR "AD") AND ("Machine Learning" OR "ML" OR "Deep Learning" OR DL" OR "artificial intelligence" OR "AI" OR "research software" OR "scientific software") AND "Hidden Technical Debt"

We tested the final search string on Google Scholar and the search returned five out of the seven control papers within the first two pages. We then executed the search in the primary libraries recommended for software engineering reviews by Kitchenham and Brereton [41], IEEE Xplore[2], ACM[3], Springer[4], and ScienceDirect[5].

*3.3.2 Inclusion and Exclusion Criteria.* The semantics of words can interfere with a search string, leading to the selection of irrelevant papers, necessitating a need for inclusion and exclusion criteria [19, 41]. Table 3 summarises the criteria we used to filter relevant studies.

We included works published in English [I1] to avoid translation issues, and restricted the time span to studies from 2010 onwards [I3] to capture the growth of TD research leading up to the emergence of AD in 2020 [2, 11]. To ensure consistency in what counted as an ML/DL system [9], we included studies that examined ML/DL components within a broader software or data-processing workflow such as training pipelines or integrated ML modules rather than papers focused solely on isolated algorithmic innovations without system-level context. However, we retained studies involving ML frameworks or experimental prototypes when they explicitly discussed implications for system behaviour such as scalability, performance, or maintainability, as these align with our definition of ML/DL systems [I4]. We also included studies that specifically addressed TD within ML/DL systems [I5], as well ML/DL-related code smell studies [I6], to avoid misclassification of AD-related smells given the niche and emerging nature of AD. Given this emergent status, both primary and secondary studies were included to contextualise AD (RQ2).

---

[2]http://ieeexplore.ieee.org
[3]http://dl.acm.org
[4]https://link.springer.com/
[5]https://www.sciencedirect.com/

Table 3. Inclusion and exclusion criteria used during the selection of primary studies to determine their eligibility for analysis.

| Code | Inclusion Criteria |
| --- | --- |
| I1 | Papers published in English |
| I2 | Journals, conference, workshop proceedings, and book chapters |
| I3 | Papers published since 2010 |
| I4 | Studies focusing on ML/DL algorithms |
| I5 | Studies focusing on TD in ML/DL |
| I6 | Studies discussing code smells in ML/DL systems |
| **Code** | **Exclusion Criteria** |
| E1 | Earlier versions of extended works (e.g., conference abstracts) that were later published as full papers (papers with full text available) |
| E2 | Duplicate papers |
| E3 | Preprints |

To avoid threats to conclusion validity [94], earlier versions of extended works later published as full papers (e.g., conference abstracts superseded by journals) were excluded from the study [E1]. Additionally, if a paper was available in multiple search libraries (as it often happens with joint ACM/IEEE conferences), only one single instance was kept [E2] to avoid duplication. We excluded preprints from the study [E3] because, while they may contain state-of-the-art information, they have not undergone peer review and may be subject to revision. Excluding them ensures that the survey includes only verified, peer-reviewed knowledge, thereby maintaining the reliability and credibility of our findings.

To ensure consistency across the different studies, we treated an ML/DL system as any software system in which a ML/DL model forms part of a functional workflow, such as data preprocessing, feature extraction, model training, inference, or deployment. We acknowledge that the included studies vary in granularity as some examine algorithms or models in isolation, some evaluate frameworks or experimental prototypes, and others analyse fully deployed systems. During screening, we included a study as long as the ML/DL component operated within a broader process or was discussed in a way that had implications for system-level behaviour (e.g., computational cost, scalability, model degradation, or design trade-offs). Our analysis does not assume that all papers adopt the same system definition. Instead, we coded AD manifestations in the context used by each study and then normalised these codes during synthesis. This allowed us to compare algorithm-level and system-level AD manifestations without forcing a uniform system boundary that the primary studies did not share.

*3.3.3 Filtering.* After executing the search query, a total of 710 papers were downloaded and managed using Zotero (i.e., a reference management software). The duplicate papers were removed and a refined set of 567 papers were left. The first author independently screened the titles and read the abstracts, excluding papers failing the inclusion and exclusion criteria (Section 3.3.2), e.g., those on Test Debt in non-ML systems, yielding 276 papers. The screening results were analysed in consultation with the TD expert in Section 3.2 through open discussion, leading to consensus decisions and thereby negating the need for inter-rater reliability measures. We then assessed the full texts of the remaining papers based on the inclusion and exclusion checklist (Section 3.3.2), resulting in a total of 25 studies.

To identify additional relevant studies, we performed forward and backward snowballing on Google Scholar [44, 52, 90], examining the citations and references of our primary studies. We found a total of 10 additional studies that met our inclusion/exclusion criteria after two iterations when no new relevant papers were identified, indicating saturation.

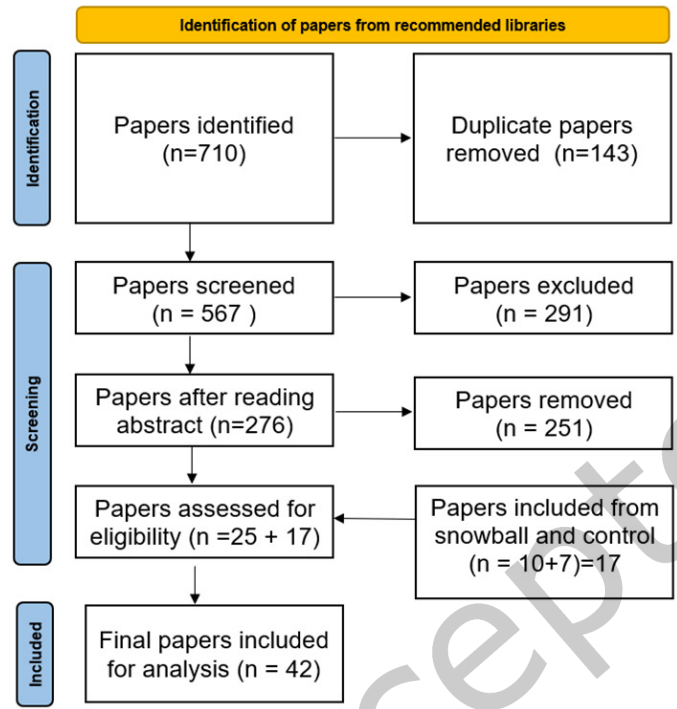

Fig. 3. PRISMA diagram illustrating the selection process of the study, adapted from Moher et al. [54].

We obtained a total of 35 papers from both the first filtering, second filtering, and snowballing processes. Considering the earlier seven control papers, we gathered a total of 42 papers (Figure 3) for our survey. Among these, seven "directly mentioned" AD (approximately 17%) and in 35 papers (83%) AD was not directly mentioned ("implied"). The paper distribution is shown in Figure 4, illustrating the control, primary, and snowballed papers.

The distribution of our primary studies based on the venues is presented in Figure 5 and the complete list is available in our online repository[1].

*3.3.4 Quality Assessment (QA).* We carried out Quality Assessment (QA) out on the selected papers to ensure rigor and relevance by evaluating their content against a checklist (Table 4) and mapping them to a ranking scale [45, pp.5]. The following QA criteria (QA1-QA4) were adapted: i) QA1 evaluated the type of work carried out to ensure the paper was based on evidence rather than opinion, (ii) QA2 assessed the relevance of the paper to any of the proposed RQs (RQ4.1–RQ3), iii) QA3 assessed the rigour of the methodology, and iv) QA4 assessed the credibility of the findings. Each criterion was rated on a three-point scale (Yes = 2, Partial = 1, No = 0). Additional QA criteria such as reflexivity were not included to avoid excluding relevant studies, given that AD is a relatively new research area. At the end of this assessment, all the included papers passed the QA.

*3.3.5 Extraction and Synthesis.* We first distinguish between data extraction and synthesis. Extraction refers to the capture of definitions, descriptions, and examples of AD-related phenomena reported in the primary studies, without interpretation. Synthesis refers to the subsequent analytical process in which the data we extracted were interpreted and abstracted to derive consolidated definitions, categories, and themes across studies.

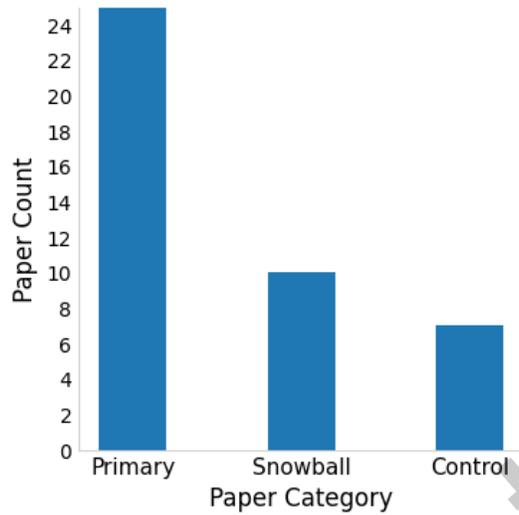

Fig. 4. Distribution of selected papers categorised as control, primary and snowballed papers.

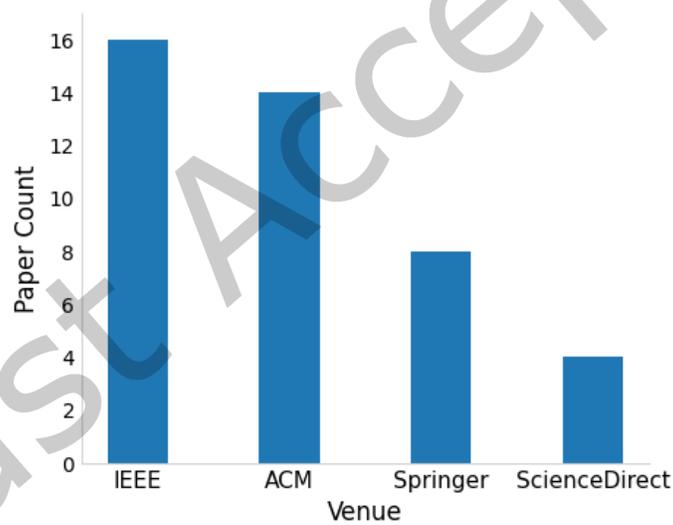

Fig. 5. Distribution of selected papers based on the venue. The others refer to the venues of the papers that we found from the snowballing process.

Table 5 presents the data extraction form, that we designed in a spreadsheet format to capture data relevant to our RQs. The first author designed the extraction form, and the same expert from Section 3.2 reviewed the fields to ensure complete data were captured. We remained flexible in adding extra columns to accommodate newly identified data categories, ensuring no relevant information was missed.

Table 4. Quality assessment criteria used to evaluate the included studies. Each criterion was rated on a three-point scale (Yes = 2, Partial = 1, No = 0) to determine eligibility for inclusion in the final analysis.

| QA | Assessment Criteria |
| --- | --- |
| QA1 | Does the study provide empirical or theoretical research findings, or is it an opinion piece? |
| QA2 | Does the study directly or indirectly address any of the RQs? |
| QA3 | Does the study employ a clearly defined methodology suitable for answering its RQs? |
| QA4 | Does the study provide clearly stated and justified findings? |

Table 5. Template employed for structured data extraction (DE) during the review process.

| Variable Code | Header | Description | RQ Answered |
| --- | --- | --- | --- |
| [DE1] | Paper Code | A unique ID / code for the paper being considered | N/A |
| [DE2] | Title | The title of the paper being considered | N/A |
| [DE3] | Author | Names of the author(s) of the paper | N/A |
| [DE4] | Year | The year of publication | N/A |
| [DE5] | Abstract | The paper abstract is copied here | N/A |
| [DE6] | AD Mention | Indicates whether AD was explicitly mentioned or implied through related discussions | RQ1 & RQ2 |
| [DE7] | Initial RQs of the Paper | RQs posed by the authors on AD or TD | RQ1 & RQ2 |
| [DE8] | Proposed Definition of AD | Direct words used by the author to define AD | RQ1 |
| [DE9] | Keywords from the Definition | Specific terminology used by the author to define AD | RQ1 |
| [DE10] | AD Indicator | Specific evidence or metrics signaling AD (e.g., performance degradation) | RQ1 |
| [DE11] | Example of Text Indicating AD | Example phrases or terms hinting at AD beyond formal definitions | RQ1 & RQ2 |
| [DE12] | Dataset Link | URL to the dataset used | RQ1 |
| [DE13] | ML/DL Smells | ML/DL-related smells identified in the study | RQ3 |

**Thematic Analysis:** We employed thematic analysis [17], to derive concepts from the included literature and supplementary datasets, addressing RQ1–RQ3. The coding process followed two phases. First, the aforementioned expert in TD (Section 3.2) developed the coding criteria based on a preliminary synthesis of relevant literature. Second, the first author completed a two-hour training session led by the the expert on how to apply the coding scheme to the source materials (literature excerpts and code comments). After the initial round of coding, the domain expert independently reviewed and verified all codes to ensure consistency and analytic rigour.

Below we provide examples to illustrate how raw comments were interpreted, coded, and mapped to final categories. To enhance transparency, we provide worked examples demonstrating how the raw text was translated into initial codes and assigned to their final high-level categories. E.g., the comment `Add direct conversion, since creating an intermediate array might be very slow` was coded as "avoid unnecessary intermediate operation" because it highlighted a performance problem. This code was mapped to our final category Inefficiency. Likewise, the comment `batchSize is fixed to one` was coded as "fix rigid batch size constraint", and assigned to the category Scalability Limitations. Similarly, the comment `Add an epsilon` was coded as "missing numerical stabilisation" and categorised under Model Degradation. Finally, the comment `TODO: sharing = false?`, was coded as "Unclear impact" and categorised as "Unclassified", since the comment appears to be a question with unclear AD impact. Without context on how the sharing parameter affects the model performance, classification is unclear. These examples demonstrate how we moved from raw text to codes and then to the final high-level categories. This is illustrated in Figure 6.

Similarly, we extracted textual evidence extracted from the included literature, treated them as raw input, and coded them using the same procedure. For example, the statement *Model Debt originates from suboptimal feature selection processes, neglected hyperparameter tuning, and inefficient deployment strategies* [16], was coded as "inefficient model design and training practices" and was mapped to the high-level category Scalability Limitations, as it describes practices that hinder efficient scaling of model training and deployment.

We adopted an expert validation method for the final categories through iterative consensus discussions rather than statistical inter-rater agreement metrics such as Cohen's Kappa [24]. This expert validation is important because our analysis required interpretive judgments about conceptual relationships across heterogeneous sources rather than straightforward categorical assignments. Also, collaborative refinement through expert consensus is an established approach in qualitative systematic reviews for synthesising themes [68]. Hence, this continuous consensus-driven validation, in which the expert verified the final codes generated by the first author, enhanced accuracy of the thematic coding.

**Methodology for RQ1: Definition of AD:** To address RQ1 (defining AD in ML/DL systems and identifying its key components), we employed a two-way approach (Figure 7), combining a review of literature with supplementary analysis of AD-labeled comments to complement the literature findings, given the limited available literature on AD. First, we synthesised definitions of AD from the subset of seven included studies that explicitly mentioned AD. These seven studies were selected because they contained explicit conceptualisations of AD, as opposed to merely implicit references to related TD phenomena. This synthesis enabled us to extract the preliminary definition of AD in ML/DL systems.

Second, to identify the key components of AD (e.g., technical impacts like efficiency, performance, scalability), we analysed the dataset of AD-labeled SATD comments from Liu et al. [49], as only seven survey papers detailed specific AD manifestations. This dataset, containing 7,159 labeled SATD instances from seven DL repositories, provided empirical examples of AD's components not fully captured in the literature definition. We applied Braun and Clarke [17] thematic analysis framework to categorise the comments based on technical impact derived from the original definition of AD by Liu et al. [49].

We followed the thematic analysis procedure described in Section 3.3.5, where the aforementioned expert developed the coding criteria and the first author performed the coding after a 20-comment pilot. The first author then thematically coded the AD samples in the dataset, which resulted into the themes: inefficiency, Scalability Limitations, Model Degradation, and Unclassified based on observed patterns in the data. To ensure reliability, the expert independently verified the final codes generated by the first author and the theme assignments.

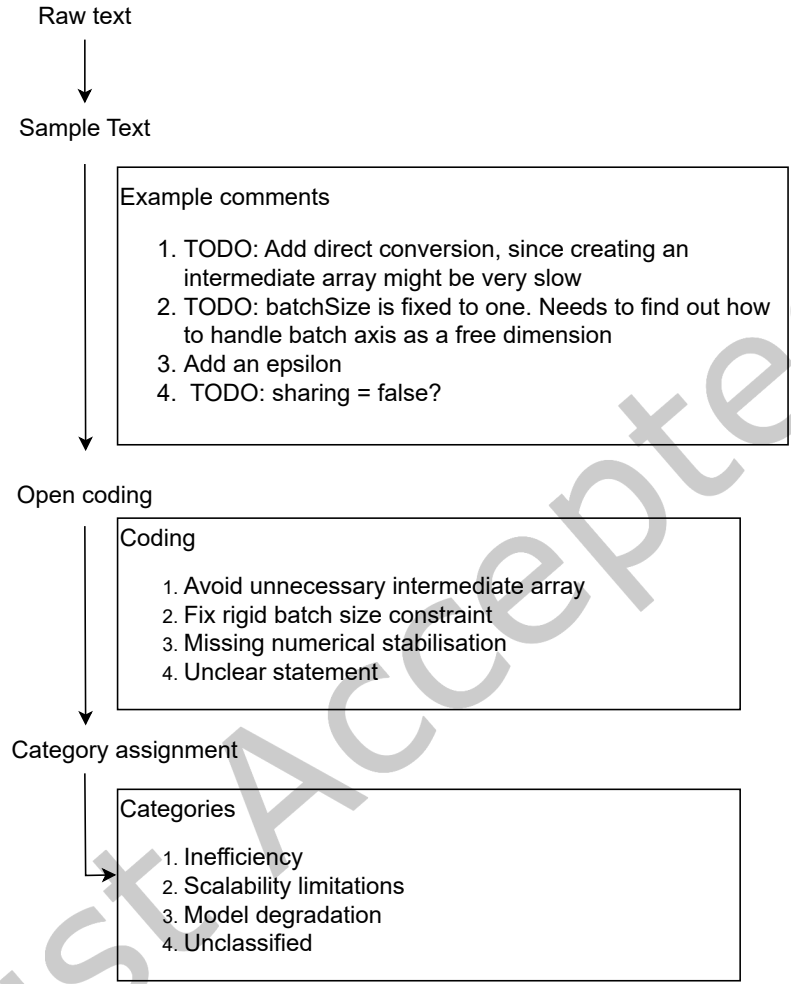

Fig. 6. Coding framework for the SATD data

Overall, this integrated approach of using the literature and dataset ensured that the literature-based definition was supported with empirical components from the dataset, providing a comprehensive understanding of AD in ML/DL systems. The entire process is illustrated in Figure 7.

**Methodology for RQ2: Implicit and Explicit AD research:** To investigate RQ2 (i.e., the extent and ways AD has been addressed in ML/DL research), we examined all included 42 studies. We reviewed papers from pre-2020 (before AD's formal definition by Liu et al. [49] and first mention) and post-2020 (after AD's formal mention) periods to capture AD-like discussions across time. From this set, we conducted full text reading, identifying papers relevant to ML/DL systems that discussed AD issues based on our inclusion/exclusion criteria. While reading these papers, we categorised them into two groups: explicit and implicit papers. The explicit papers directly mentioned AD as a named concept, while the implicit papers discussed AD-related challenges, such as data drift or suboptimal model architectures, without explicitly mentioning AD, often framing these issues within broader TD categories like Data Debt and Model Debt [16, 73]. This categorisation allowed us to map the literature to the empirical findings of this study, providing a comprehensive understanding of AD research in the literature.

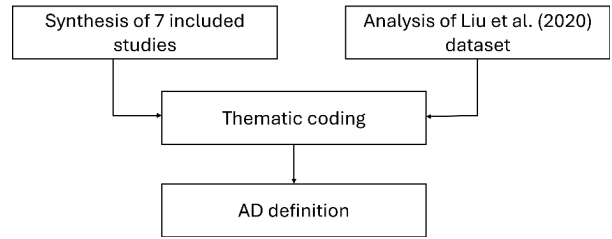

Fig. 7. RQ1 Methodology, showing the literature review complemented by analysis of Liu et al. [49]'s dataset to derive the definition for AD

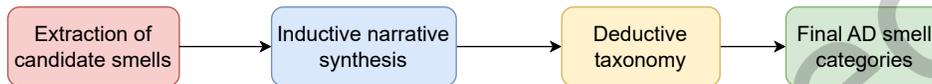

Fig. 8. Methodology for RQ3, illustrating the steps we followed to reclassify code smells as AD items

**Methodology for RQ3: AD smells in ML/DL Systems:** To address RQ3, identifying the main categories of AD smells in ML/DL systems, we developed a three-phase methodology (illustrated in Figure 8) to synthesise and categorise AD-related smells from the 42 included studies. We defined AD smells here as literature-reported indicators of suboptimal algorithmic choices, leading to model degradation and poor scalability based on the definition of AD in RQ1.

First, we extracted all ML/DL-specific TD and code smells reported in the primary studies. We included smells meeting the following criteria derived from the definition of AD: i) algorithmic relevance (e.g., inefficiencies due to algorithm choice or configuration), ii) performance-related impact (e.g., accuracy degradation), and iii) scalability-related impact (e.g., memory bottlenecks, difficulties with large datasets). Smells were excluded if their primary impact was structural, organisational, or stylistic (e.g., long methods, naming issues) rather than algorithmic behaviour or ML-specific performance, scalability, or degradation concerns.

Second, we conducted an inductive narrative synthesis [68] to derive preliminary AD smell groupings directly from the extracted data, without relying on pre-existing classifications. Each smell's description, cause, and impact were compared across studies and grouped into thematic clusters (e.g., inefficient algorithmic operations, hyperparameter misconfiguration). E.g., a situation in ML where the computed statistics or metrics used for model training do not accurately represent the underlying population or data distribution, the smell was categorised as *non-representative statistics estimation*. This stage reflects a data-driven approach and does not rely on pre-existing taxonomies.

Third, we performed a deductive classification step, in which the inductively derived smell groups were mapped onto a broader, theory-informed taxonomy adapted from Pepe et al. [67]. Each group that was inductively derived was mapped to broader categories (data-related, algorithm-design-related, ML-knowledge-related). E.g., *plain* *old data, non-representative statistics, out of domain data, and unstable data dependencies* were all mapped to data. The first author and the TD expert (Section 3.2) reviewed categorisations, discussed ambiguous cases, and reached consensus on final classifications. E.g., ML smells such as *long method* was removed from the list as it is more related to Design smells rather than AD smells. This two-stage process which involved inductive grouping followed by deductive alignment, ensured that AD smells emerged from the literature while remaining comparable with established ML/DL smell taxonomies.

## 4 Results

### 4.1 RQ1: Definition of AD

We read and analysed the primary studies explicitly mentioning AD, and found a repetition of Liu et al. [49] definition for AD, often with scarce tweaks [50, 74, 86], while emphasising its impact on performance. However, these studies rarely specified which performance aspects are affected. Table 6 highlights key sentences that we extracted from the definitions of AD.

Table 6. Key sentences extracted from the definitions of AD used in "direct mention" papers.

| # | Keywords | Reference |
|---|---|---|
| 1 | Sub-optimal implementations of algorithm logic | [49, 84] |
| 2 | In the DL frameworks | [8, 49, 50, 74, 85] |
| 3 | Pull down the system performance | [74, 85] |

While these definitions established a conceptual foundation for AD, they lacked empirical detail on the specific components and real-world manifestations of AD. Building on these definitions, and using the methodology in Section 3, we analysed comments labelled as AD from Liu et al. [49]'s dataset to identify key components. This was done to complement the review of papers by grounding the scope of AD in real-world developer experiences. We identified three broad categories through open coding. The categories were: inefficiency, scalability limitations, and degradation, assigning each comment a primary category based on its dominant technical challenge, while noting secondary impacts where relevant (e.g., inefficiency due to lack of scalability). These categories aligned with common software quality concerns. The frequency distribution of the final AD categories and representative examples of their classification is shown in Table 7. This table covers all four categories across the coded AD instances, providing an overview of their prevalence. Detailed examples of each AD category, along with the rationale for their classification, are provided in our online repository[1].

To further refine our understanding and link these categories to concrete technical problems, we subsequently identified specific *issues* for each comment, which describe the precise AD problem (e.g., suboptimal algorithms, insufficient training data) and mapped them to the emergent categories, as shown in Table 7. These issues, validated by the TD expert, reflected the reported challenges in ML/DL systems and were linked to software quality concerns.

An analysis of the comments within each primary category provides highlights how these issues manifest in practice, solidifying the three themes identified through the open coding process. *Inefficiency*, the first challenge, arises from redundant computations or suboptimal algorithmic logic, resulting in increased time and memory usage. E.g., a comment from the dataset `We will simply do a two-pass. More efficient solutions can be written, but I'll keep the code simple for now.` illustrates this point. The phrase "more efficient solutions can be written" explicitly acknowledges that the current method is not optimal but is being used as a temporary fix, possibly for simplicity; temporary fixes could increase overhead, a characteristic of TD. Similarly, this comment `A slightly inefficient way to get one-hot vectors, but fine for low vocab (like char-lm).` highlights this. The comment states that the method is inefficient, implying that it may experience a memory bottleneck for larger vocabularies but is tolerable for smaller datasets. These examples highlight how inefficiency manifests in non-optimised designs that inflate computational costs, particularly when scaling to larger datasets.

Table 7. Sample AD comments from Liu et al. [49]'s dataset

| Category | Example Comments | Percentage |
|---|---|---|
| Inefficiency | • `TODO (jeff, opensource): This should really be a more interesting computation. Maybe turn this into an MNIST model instead.`<br>• `A slightly inefficient way to get one-hot vectors, but fine for low vocab (like char-lm).` | 31% |
| Scalability Limitations | • `The current implementation is O(N²), it should be possible to do this in O(N logN).`<br>• `AD HOC: This assumes that the worker has a hostname, which is not the case for MPI workers.` | 14% |
| Model Degradation | • `Not enough data to do much more than this. stop(All values must be positive).`<br>• `Fix for lost factor information, may not be needed?` | 31% |
| Unclassified | • TODO: sharing = false?<br>• TODO: Conflict of parameter order: filter_shape or num_filters first? | 24% |

*Scalability limitations*, which is the second dimension, relates to algorithmic and ML pipeline implementation. Scalability limitations arise when algorithmic or ML pipeline designs restrict performance on larger datasets or complex models. The causes include inefficient resource utilisation, suboptimal feature selection, neglected hyperparameter tuning, or poor deployment strategies. E.g., the comment `AD HOC: This assumes that the worker has a hostname, which is not the case for MPI workers`, highlights a limitation in distributed computing. This is likely because certain assumptions in the algorithm do not hold for Message Passing Interface workers used in high-performance computing environments, hindering parallel execution in large-scale ML/DL systems. In other words, the algorithm's design leads to AD (e.g., simplifying assumptions for local testing), creating hidden constraints that limit performance when deployed at a large scale. This exemplifies how AD can impact scalability, particularly in distributed or high-performance ML/DL environments, where architectural assumptions no longer hold.

The third dimension, *Model degradation*, encompasses issues such as numerical instability, overfitting, concept drift, and data integrity loss, which compromise E.g., the comment `Not enough data to do much more than this. Stop ("All values must be positive")` indicates a situation where insufficient data could lead to underfitting, which degrades model performance. Similarly, `Fix for lost factor information, may not be needed?` suggests potential data integrity issues that could reduce computational accuracy. These instances demonstrate how AD contributes to performance decline through algorithmic or implementation flaws.

In summary, our analysis allowed us to characterise the scope and impact of AD. We found that AD as initially defined by Liu et al. [49], centers on performance issues due to algorithmic flaws, yet lacks detail on specific the impacts. The thematic analysis of the AD comments from [49] dataset clarified these impacts, identifying inefficiency, scalability limitations, and degradation as recurring themes across ML/DL contexts. These categories, prevalent across ML/DL systems highlight the need for a refined definition that captures these dimensions.

> **RQ1 Answer — AD Definition.**
> AD is a TD type that arises when algorithmic choices are implemented in ways that lead to limited scalability and model degradation relative to system goals. AD manifests in three key forms: i) inefficiency, where redundant computations or non-optimised logic increase computational costs; ii) scalability limitations, where algorithms struggle with large-scale processing or distributed execution; and iii) degradation, where performance decline due to specific algorithmic flaws or implementation.

### 4.2 RQ2: Research on AD in ML/DL Systems

We explored whether prior research in ML/DL systems, addressed AD-related issues without using the term AD. For this research, an "implicit" mention of AD is defined as a discussion regarding any of these categories (i.e., that we identified from RQ1) for ML/DL systems: i) algorithmic inefficiency (use of an algorithm when a more efficient one exists, unnecessary computations within an algorithm), ii) scalability issues related to algorithmic implementation (algorithms with high time or space complexity that limit scalability with larger datasets or more complex models), and iii) degradation challenges related to complex algorithmic logic (when algorithms or models fail to perform as expected due to implementation issues).

To empirically validate this hypothesis and identify the presence of implicitly discussed AD in the literature, we conducted a thematic analysis of the selected studies. Table 8, provides sample text from selected papers (with a reference to the paper that is being referenced for each specific quote), the AD-issue that the example discussed, and an explanation of why we categorised it as AD.

A closer examination of these examples clarifies how challenges that do not explicitly use the term AD align with our debt categories, beginning with model degradation issues. E.g., Sculley et al. [73] discussed issues related to model degradation in ID1. "Unstable data dependencies increases the computational cost of retraining models due to the model degradation". The text aligns with our definition of AD because unstable data dependencies can lead to unpredictable model performance, a key aspect of model degradation. A mis-calibrated input signal makes an ML/DL model to learn flawed patterns, leading to model degradation when a correction is applied. This scenario reflects AD as the model lacks robustness to data changes, indicating poor design or implementation. The tight coupling to flawed data could reveal a lack of modularity, while potential difficulties in identifying and correcting mis-calibrations increase maintenance costs. If the algorithm was designed without considering potential data shifts or noise, leading to significant performance decline upon data correction, this highlights an algorithmic flaw. Ultimately, the necessary model retraining due to these effects highlights the direct impact of AD, where insufficient foresight during development results in degraded performance and increased operational overhead.

Similarly, the literature reveals discussion of algorithmic inefficiency, a second key facet of AD, as highlighted by examples in ID2. In ID2, OBrien et al. [64] stated that "ML software carries plenty of unique challenges, that could lead to inefficiencies during algorithm implementation. Uneducated solutions by "unaware developers" may have to be revisited.", raising issues on the inefficiency caused as a result of lack of domain knowledge which could lead to inefficiency on how to correctly implement a component. These uncertainties and unresolved questions in the development process, e.g., "TODO: Something smarter?" and "TODO: do we want gainsbiases to be trainable?", directly contributes to AD. Known algorithmic inefficiencies, represented by the need for a "smarter" implementation, become embedded within the system. Ambiguous design decisions, such as whether gains and biases should be trainable, lead to potential TD through increased complexity, maintainability challenges, and hindered scalability. These unresolved issues represent compromises made during development, which accumulate over time, creating a debt that requires future refactoring or redesign to address, thus illustrating the accumulation of AD.

Table 8. Challenges associated with AD and their supporting evidence

| S/N | Paper Citation | #paper | AD-Issue | Sample Quote(s) | Explanation |
|---|---|---|---|---|---|
| ID1 | Bogner et al. [16], Giray [32], Habibullah et al. [35], Li et al. [47], Liu et al. [48, 49], Nikanjam and Khomh [62], OBrien et al. [64], Sculley et al. [73], Vélez et al. [82], Villamizar et al. [87], Wang et al. [88] | 12 | Model degradation | "For example, consider the case in which an input signal was previously mis-calibrated. The model consuming it likely fit to these mis-calibrations, and a silent update that corrects the signal will have sudden ramifications for the model" Sculley et al. [73] | A mis-calibrated input causes a model to overfit or bias, leading to degradation upon correction. |
| ID2 | Biswas and Rajan [15], Bogner et al. [16], Chaudhary et al. [22], Giray [32], Humbatova et al. [36], Jia et al. [38], Jiang et al. [39], Li et al. [47], Liu et al. [48, 50], Nahar et al. [60], OBrien et al. [64], Paleyes et al. [65], Pepe et al. [67], Sun et al. [79], Vélez et al. [82], Villamizar et al. [87], Wang et al. [88], Zhang et al. [93] | 19 | Inefficiency | "Doubts on current design decisions can be due to questioning the qualities of a current implementation (TODO: Something smarter?) or questioning the design decisions in place (TODO: do we want gainsbiases to be trainable?)" Bogner et al. [16] | Unresolved design choices delay efficient algorithms, raising runtime costs. |
| ID3 | Amrit and Narayanappa [6], Biswas and Rajan [15], Cruickshank and Kohtz [26], Humbatova et al. [36], Jiang et al. [39], Li et al. [47], Liu et al. [49], Munappy et al. [58], Paleyes et al. [65], Pepe et al. [67], Vélez et al. [82], Vidoni [85], Villamizar et al. [87, 87], Zhang et al. [93] | 16 | Scalability Limitations | "This AI-specific debt type regards suboptimal practices in the design, training, and management of AI models. Most prominently, Model Debt originates from suboptimal feature selection processes, neglected hyperparameter tuning, and model deployment strategies that are inefficiently designed" [16] | Suboptimal feature selection and tuning increase complexity, limiting scalability with larger datasets. |

The final category, scalability limitations, is also implicitly addressed in the literature, particularly when discussing suboptimal model design practices that hinder large-scale processing (ID3). In the third quote from the table (ID3), post-2020, Bogner et al. [16] discussed suboptimal practices in the design, training, and management of models leading to model degradation. They highlighted the impact of suboptimal practices, noting how these issues manifest as deficiencies that are related to model components. Specifically, they highlighted that *"Model Debt originates from suboptimal feature selection processes, neglected hyperparameter tuning, and poorly engineered model deployment strategies"*. Inefficient feature selection in ML/DL training directly impacts scalability by increasing data dimensionality, leading to the "curse of dimensionality" and data sparsity, which demand more training data and computational resources. If the algorithm that performs the hyperparameter tuning is inefficient, or gets stuck in local minima, then that becomes AD. This results in higher computational costs and longer training times.

Additionally, poorly optimised hyperparameters may lead to suboptimal model performance, potentially including underfitting or overfitting, if proper regularisation techniques are not implemented. Furthermore, larger feature sets require increased data storage and processing power, creating bottlenecks in data pipelines. Consequently, effective feature selection is essential for managing computational demands and ensuring the scalability of ML/DL models, as it reduces dimensionality, decreases costs, and improves generalisation.

While these findings confirm that the underlying issues of AD are prevalent in ML/DL research, they also necessitate a clear distiction from other closely related TD types. These findings highlight that while AD might share some characteristics with TD types like Model Debt, it is distinct in that it specifically addresses inefficiencies in algorithmic logic. However, unlike TD types such as Model Debt, which encompasses a broader set of issues such as model deployment, AD focuses solely on the algorithmic choices that affect scalability, efficiency, and long-term system maintainability. To distinguish AD from these TD types, we focused on how algorithmic flaws could cause problems, rather than simply what problems exist (e.g., suboptimal features).

Given the potential for overlap and misclassification, especially since AD indicators, such as suboptimal feature selection and unstable data dependencies, were often categorised as Data Debt or Model Debt, we used a checklist derived from our RQ1 definition of AD. This checklist ensured that only algorithmic inefficiencies, scalability challenges, and degradation issues directly caused by suboptimal algorithmic logic were categorised as AD. E.g., unstable data dependencies could impact both Data Debt and AD. However, when instability was caused by algorithmic logic failing to adapt to data shifts, it was classified as AD. Similarly, feature selection issues were categorised as AD when they resulted from inefficient algorithmic implementation rather than improper data preprocessing.

Applying this classification criteria, we found a total of 83% (approximately 35/42) of the papers (which are available in the supplementary material[1]) discussed AD-related issues implicitly (illustrated in Figure 9). This suggested that prior research had indeed investigated issues related to AD, even before the formal definition of the term, indicating a growing concerns of these issues in the ML/DL community. The lack of mention of AD is likely due to the fact that these papers were published prior to the first known mention of AD in 2020 or during a period when the work that uncovered AD might have been under review. Another reason might be that researchers were focusing on broader TD types without considering AD or AD-related issues were being misclassified under other TD categories (e.g., Model Debt or Data Debt) for papers published after AD was formally mentioned. However, the causes of these misclassifcations is out of the scope of our survey.

**RQ2 Summary** — Prior research implicitly discussed AD-related issues in ML/DL systems, even before its formal definition with 83% (35/42) of papers through inefficiencies, scalability challenges, or degradation. Our analysis of the primary studies revealed that AD shares characteristics with other TD types, such as Model and Data Debt, but is distinct in its focus on algorithmic inefficiencies, scalability issues, and degradation challenges.

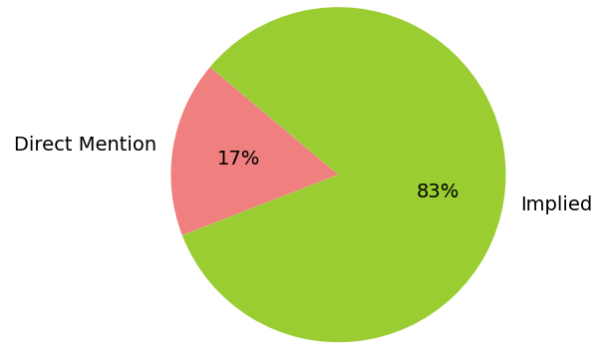

Fig. 9. Proportion of papers that explicitly mention AD Compared to those that refer to it implicitly through related concepts.

### 4.3 RQ3: Causes of AD in ML/DL systems

Code smells have long been recognised as indicators of TD, providing early warnings of potential system inefficiencies [10, 63]. However, prior studies have not explicitly examined AD smells. Our work bridges this gap by proposing a structured classification to identify and categorise AD smells. Using thematic analysis, we reclassified ML/DL-specific smells found in prior studies to highlight their connection to AD.

Figure 10 shows the reclassification of code smells as AD items. It illustrates how smells (like 'bias with batch norm') were reassigned from general ML TD to AD based on their algorithmic impact. As these papers did not explicitly discuss AD, researchers may have classified these smells under different TD types. Nevertheless, these smells serve as a valuable resource for analysing AD, effectively opening avenues for future work.

Using the methodology in Section 3, we identified and categorised nine AD smells, which we describe and illustrate in Table 9. The smells include *plain old data, non-representative statistics, bias with batch norm, unstable data dependencies, out of domain data, no scaling before sensitive oeration, unormalised feature, inadequate ML expertise, and hyperparameters* [16, 62, 73, 81, 93]. The classification process involved reviewing and then reassigning ML/DL smells as AD smells after an analysis of their underlying causes and implications. We briefly present these smells.

**Plain old data** refers to unformatted, text files that contain human-readable content that typically does not contain any special formatting or markup language [64, 73, 88]. In ML systems, complex computational representations of features, weights, and model parameters are represented using basic data types such as floats and integers. In many cases, these raw data types lack the context needed to fully understand their meaning and how they should be interpreted within the broader context of the ML system. This leads to AD because the system must implement complex and potentially inefficient logic (e.g., extra steps) to infer context, raising runtime costs. E.g., an image classification model that receives image data as a raw byte array, without metadata about image dimensions or color channels, would need to implement extra logic to interpret the data, introducing AD by increasing run time and complexity due to the inefficiency in memory usage.

**Non-representative statistics estimation** refers to the situation in ML where the computed statistics or metrics used for model training do not accurately represent the underlying population or data distribution [62]. When non-representative statistics are used for model training, it can introduce bias and inaccuracies into the learning process, reducing accuracy. This AD may arise from using skewed sampling that do not properly account for data characteristics or handle non-representative statistics. An example is using a biased mean calculation in a clustering model. This can lead to suboptimal model performance, reduced generalisation ability, and unreliable predictions from a small sample may group data incorrectly, leading to unreliable outputs and rework.

Table 9. Identified AD smells and their primary studies

| Smell | Primary Study | #Paper | Category |
|---|---|---|---|
| Plain old data | OBrien et al. [64], Sculley et al. [73], Wang et al. [88] | 3 | Data |
| Non-representative statistics estimation | Nikanjam and Khomh [62] | 1 | Algorithm design |
| Bias with batch norm | Li et al. [47], Nikanjam and Khomh [62] | 2 | Algorithm design |
| Unstable data dependencies | Breck et al. [18], Giray [32], Sculley et al. [73], Wang et al. [88] | 4 | Data |
| Out of domain data | Cruickshank and Kohtz [26], Liugi [51], Munappy et al. [58], Sculley et al. [73] | 4 | Data |
| No scaling before scaling-sensitive operation | Biswas and Rajan [15], Bogner et al. [16], Humbatova et al. [36], Jia et al. [38], Jiang et al. [39], OBrien et al. [64] | 6 | Data |
| Unnormalised feature | Biswas and Rajan [15], Bogner et al. [16], Chaudhary et al. [22], Humbatova et al. [36] | 4 | Data |
| Inadequate ML Knowledge | Chaudhary et al. [22], Giray [32], Liu et al. [50], Nahar et al. [60], OBrien et al. [64], Pepe et al. [67], Sun et al. [79] | 7 | ML expertise |
| Hyperparameters | Biswas and Rajan [15], Humbatova et al. [36], Jiang et al. [39], Li et al. [47], Paleyes et al. [65], Pepe et al. [67], Vélez et al. [82], Villamizar et al. [87], Zhang et al. [93] | 9 | Algorithm design |

**Bias with batch norm** Batch norm is a technique commonly used in DL models, to normalise the activation of a neural network layer [47, 62]. In DL models, it is generally beneficial to include bias terms in the learning layers with different initialisations to to stabilise, accelerate the training, and improve generalisation. The impact of batch norms on the system is usually reduced when a bias is introduced. If the batch norm is applied after dropout, it may calculate global statistics that are not representative of the actual data distribution. This results in AD due to the incorrect normalisation logic which disrupts learning (algorithmic nature), degrading model accuracy (performance). E.g., a CNN applying batch norm post-dropout may produce inconsistent predictions, where scaling larger networks would require redesign.

**Unstable data dependencies** refer to where certain signals used as input features in a system's processing exhibit qualitative or quantitative changes over time due to the dynamic nature of ML models [18, 32, 73, 88]. Integrating data that originate from external systems used as input features for the model can offer initial benefits in terms of development efficiency. However, these evolving inputs cause covariate shift on the associated ML components. This could lead to concept and data drift where the deployed model will lose its predictive capabilities [25]. This AD arises from an algorithmic flaw in the logic that results in the algorithm's inability to handle evolving inputs and degrades predictive accuracy, necessitating model retraining as input sources evolve.

**Out of domain data** ML models, naturally learn patterns that are useful to a certain task from the data that they are presented with. Performance issues could therefore arise if the data presented to these models when in use is different than the data it was trained on [26, 58, 73]. An ML model that is trained during development to detect objects from a ground perspective, may fail when deployed in an environment different from its training data (e.g., urban-trained models may fail rural inputs, requiring retraining) [51]. This generates AD via an algorithmic limitation where the model cannot generalise beyond its training scope, lowering accuracy on new inputs, and necessitating retraining as data changes across domains.

**Improper Feature Scaling** refers to the process of not transforming input features to a consistent range of features to ensure that they are on a similar scale [15, 16, 36, 38, 39, 64]. Scaling is important because it helps prevent certain features from dominating the learning algorithm due to their larger magnitude. It also facilitates convergence and improves the performance of many ML algorithms, particularly those that rely on distance-based calculations. Scalling usually focused on the algorithm's sensitivity. Failing to scale features introduces AD due to a flawed algorithmic implementation that lacks necessary preprocessing steps, allowing dominant features to interfere with learning, hinder convergence, and degrade model performance. A gradient descent model with unscaled features may converge slowly. E.g., unscaled inputs delay optimisation, requiring manual fixes.

**Unnormalised feature** In ML, an unnormalised feature refers to a feature that has not been scaled or transformed to a specific range or distribution. It means that the original values of the feature are retained without any adjustments or normalisation applied [15, 16, 22, 36]. Anomalies in the data can manifest as unnormalised features, where the values display significant variance or include a few extreme outliers. E.g., a regression model with an unnormalised feature like house size (square feet) alongside small-scale features may overfit to outliers, leading to biased predictions. These unnormalised features can introduce AD due to a flawed algorithmic implementation that lacks necessary preprocessing steps for feature normalisation, impacting the ML/DL model's ability to accurately capture patterns and make reliable inferences, reducing prediction accuracy.

**Inadequate ML/DL Expertise** ML/DL software has unique challenges [73]. Code written by developers lacking expertise and awareness of the system context may require later revision [22, 32, 60, 64, 67, 79]. Developing ML/DL software requires specialised knowledge, and deficiencies in such expertise among developers could lead to AD. This introduces AD through poor technique selection which undermines system design (algorithmic nature), reduces model effectiveness, and hinders future scaling due to inefficient implementation. Poor understanding of trade-offs in model complexity, explainability, or hyperparameter tuning can lead to "quick fixes" that accumulate AD, which later inhibits scalability and maintainability. E.g., using linear regression for a task requiring a classificatiion model, may yield poor results necessitating redesign.

**Hyperparameters** Hyperparameters refer to the configuration settings or parameters of a model that are not learned from the data but are set by the user before training the model [15, 36, 39, 47, 65, 67, 82, 87, 93]. AD can arise from inadequate hyperparameter tuning, as default values often fail to optimise model performance, leading to local optima and requiring iterative experimentation for proper configuration. For example, using a default learning rate of 0.01 for a complex DL model may lead to slow convergence. This is often caused by a flawed algorithmic implementation of the hyperparameter search process, introducing AD resulting in inefficient exploration of the configuration space. To ensure optimal model performance, it is necessary to carefully select and tune the hyperparameters based on the characteristics of the data and the problem to be solved.

After identifying these nine smells, we then broadly grouped them using Pepe et al. [67]'s taxonomy into Data, Algorithm Design, and ML Knowledge as the broad categories that contribute to model degradation, inefficiency, and poor scalability (from RQ1). Figure 10 illustrates this classification of ML/DL smells as AD smells. We note, however, that further research should be done to validate these identified smells as indicators of AD and how they overlap with other TD types.

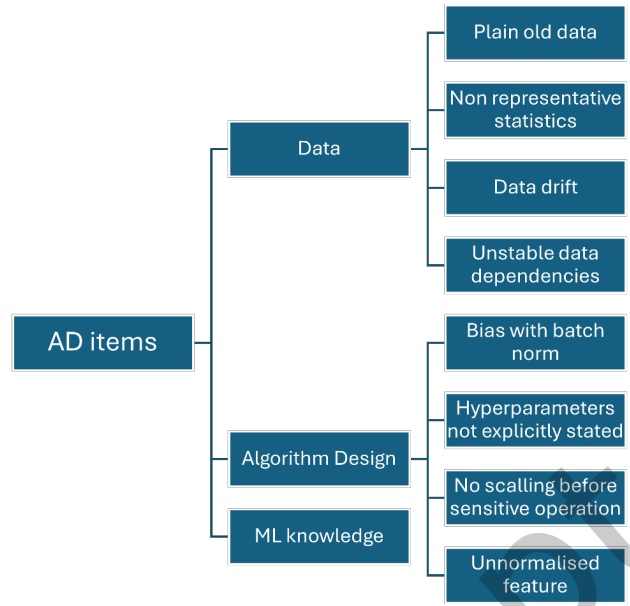

Fig. 10. Classification of ML/DL code smells as AD items based on their characteristics.

**RQ3 Answer — Causes of AD.** We identified nine smells as key causes of AD in ML/DL systems: plain old data, non-representative statistics estimation, bias with batch norm, unstable data dependencies, out of domain data, no scaling before scaling-sensitive operation, unnormalised feature, inadequate ML expertise, and hyperparameters. These stem from three broad categories data issues (e.g., poor preprocessing, drift), algorithm design flaws (e.g., misapplied techniques), and expertise gaps.

## 5 Discussion

We synthesise our findings across three dimensions: a consolidated definition of AD grounded in model degradation and inefficiency in ML/DL systems, evidence of the pervasive yet implicit nature of AD in 83% of prior work (RQ2), and identification of nine AD smells across data, algorithm design, and ML knowledge (RQ3) (Section 5.1). Next, we situate our work within the broader TD landscape, demonstrating how AD is different from TD types such as Code, Data, and Model Debt through its algorithmic focus and unique causes.

### 5.1 Principal Findings

Our literature survey and dataset analysis both converge on the interpretation that AD is a distinct TD type centred on algorithmic logic and design inefficiencies. While the included studies show that AD is primarily discussed in terms of model degradation (20/42 studies), inefficiency (20/42), and scalability limitations (12/42), our dataset analysis independently surfaced the same three patterns as the dominant categories across real-world code comments. The independent emergence of these patterns across both literature and practice as combined evidence validates AD as a distinct TD type that requires algorithm-specific mitigation strategies.

## 5.2 Synthesis of Findings

The triangulation of the literature and the analysis of the dataset strengthens the definition produced in RQ1. *AD is a TD type that arises when algorithmic choices are implemented in ways that constrain the ability of an ML/DL system to change or substitute components relative to the stated system goals.* Importantly, the analysis of the dataset not only confirms the conceptual patterns reported in the literature but provides developer-generated evidence that model degradation and inefficiency are encountered in practice. For example, comments regarding missing numerical stabilisation `Add direct conversion, since creating an intermediate array might be very slow` maps onto the theoretical construct of inefficiency identified in the review of the included studies. Second, the integration shows alignment between the literature and the dataset. While the included studies provide the conceptual vocabulary for distinguishing AD from other TD types, the dataset shows evidence of how practitioners express AD in codebases and validates that these issues are not merely theoretical. Where the literature emphasises AD as a design-level concern, the dataset analysis shows that developers frequently encounter AD at the implementation level, such as inefficient memory usage or non-scalable design patterns. This integration clarifies how unique AD is as a TD type in ML/DL systems. Both the literature and the dataset highlight AD as distinct not because of its effects alone, but because of its origins: AD emerges from algorithmic decision-making rather than from data, coding style, or software infrastructure issues alone. This positions AD as multifacet TD type. Taken together, these observations raise an important question about the nature of AD: not only where it appears, but why it emerges in the first place, a question addressed by examining developers' intentional trade-offs.

AD emerges from a trade-off, where developers choose short-term simplicity and speed over long-term efficiency. Comments (e.g., `more efficient solutions can be written` or those signaling computational overhead) reveal their awareness of the current solution being suboptimal, but still make the choice anyway at the point of decision-making. This consciousness establishes AD as an intentional phenomenon that requires organisational management, not merely technical fixes. This suggests that preventing future AD requires addressing time pressure, resource scarcity, and other constraints that drive such choices rather than improving developer awareness alone.

However, while developers may intentionally introduce AD, the research community does not always explicitly recognise it, creating a disconnect between practice and theory. We found that AD is pervasive, yet it is implicitly discussed by researchers. RQ2 revealed an asymmetry, where approximately 83% of the included studies (35/42) discussed AD-related issues without explicitly naming "AD." Examples include Sculley et al. [73]'s observation that "mis-calibrated inputs lead to model degradation" and Bogner et al. [16]'s findings that "suboptimal feature selection" leads to scalability challenges. These both align with our RQ1 definition of AD but were classified under other TD types such as Model or Design issues. This reveals a conceptual gap: practitioners and researchers encounter AD regularly but do not recognise it as a unique phenomenon. This dispersion across literature under different names, sometimes as operational concerns, design flaws, and sometimes buried within broader Model or Data Debt indicates that AD lacks a stable theoretical framing, making it difficult to compare findings, develop knowledge, or targeted mitigation strategies. Interestingly, this pattern AD across literature persists pre-2020 and post-2020, suggesting that AD's theoretical and implicit nature is not a recent phenomenon but reflects a long-standing gap in TD research.

To understand how this implicit recognition translates into concrete technical symptoms, we examine the specific AD smells we identified. We identified nine AD smells across data, algorithm design, and ML knowledge categories. These smells indicate insufficient data management practices that cascade into algorithmic problems and developers' lack of specialised knowledge to make optimal algorithmic decisions. Notably, algorithm design and ML knowledge categories represent the core of AD, which are the algorithmic logic itself and the expertise to implement it optimally. Data-related smells, while present, often amplify AD rather than cause it directly (e.g., unstable data makes algorithmic inefficiency worse). This finding supports our argument of positioning AD as distinct TD type from Data Debt, which focuses on data collection, labeling, and preprocessing independently of algorithm choice.

### 5.3 Positioning AD Within TD Literature

To situate our results within existing knowledge, we map how our findings extend the current conceptualisation of TD in ML/DL systems. Our findings build upon and extend prior TD research. The three components of AD that we identified which include scalability limitations, model degradation, and inefficiency, extends prior definition of AD by Liu et al. [49], narrowed down to performance. Our consolidated definition incorporates a broader manifestation triple, and not only runtime inefficiency, thus better capturing the systemic nature of AD in ML/DL systems relative to system goals.

Building on the expanded characterisation of AD, the next step is to examine how it reinforces and operationalises insights from foundational TD research. AD confirms Sculley et al. [73]'s foundational observation that ML systems incur hidden forms of TD beyond traditional Code Debt. Specifically, our RQ1 definition of AD and RQ3 smells validate the implicit recognition that issues such as data dependencies, feature engineering, and model training choices accumulate into long-term inefficiencies such as model degradation and poor scalability. By providing an explicit taxonomy, our work operationalises what [73] stated but did not fully characterise.

Extending beyond theoretical alignment, our findings also reveal how frequently AD-related issues surface in the empirical literature, even when not labelled as AD. Our RQ2 finding that 83% of papers implicitly discuss AD-related issues confirms our key hypothesis of AD existing in practice but lacking theoretical backing. This implicit prevalence suggests AD is not a marginal phenomenon but a central feature of ML/DL development. This recognition of AD as a pervasive challenge implies that practitioners must be proactive to address AD while developing ML/DL systems. While our taxonomy provides a first step, future work should investigate how AD accumulates across different contexts and how mitigation strategies can be operationalised in practice. Given this widespread yet implicit presence, it is important to contextualise AD within the broader TD landscape and how it differs from other TD types.

### 5.4 AD and Other TD Types

Our findings position AD as a distinct, yet interdependent, form of TD in ML/DL systems. While [16, 37, 73] recognised multiple forms of TD in ML systems, none explicitly studied AD as a distinct type to investigate its causes and manifestations. This work positions AD as occupying a unique niche in the TD landscape. While Code Debt stems from structural flaws (e.g., poor naming convention, excessive nesting, and field duplication) and with a focus on maintainability, Data Debt arises from data collection, labeling, or preprocessing issues, with a focus on data quality and governance. Likewise, Model Debt emerges from model and model-achitecture selection or their integration choices in system production. AD arises from suboptimal algorithmic logic and insufficient expertise at the point of decision-making, with effects on model scalability. This means that AD accumulates at the logic level, in the algorithms themselves, independent of clean code practices. This positioning has implications: misclassifying AD as Code Debt (e.g., refactoring code structure) or Data Debt (e.g., cleaning data) will address symptoms but fail to address the root cause which is suboptimal algorithmic logic.

Importantly, AD does not exist in isolation but often overlaps with and amplifies other TD types. E.g., inefficient algorithmic choices can exacerbate Data Debt by increasing sensitivity to noisy or drifting data, while overly complex optimisation strategies may intensify Model Debt by complicating integration and maintenance. These interactions help explain why AD is frequently misclassified in practice, where its symptoms often surface as performance regressions or maintainability issues that resemble Data or Model Debt, even though the root cause lies in algorithmic design decisions.

This distinction has practical and conceptual implications. Addressing AD through conventional remedies (e.g., by refactoring code structure or improving data quality) may alleviate surface-level issues without resolving the underlying algorithmic inefficiencies. As a result, AD can persist and continue to accumulate, despite apparent improvements elsewhere in the system. Recognising AD as a separate, logic-level form of debt therefore enables more targeted mitigation strategies and clarifies why existing TD management approaches are often insufficient for ML/DL systems. These overlaps and dependencies help explain why AD is often difficult to detect and manage in practice, and motivate the need for both practitioner-facing interventions and targeted research directions discussed next.

### 5.5 Addressing AD

To understand why AD remains largely implicit in the literature despite its prevalence, we first examine how terminological and disciplinary factors shape the way AD-related issues are framed. Given that the term is recent [49], prior work was unable to reference this terminology, discussed the same phenomena under different concepts such as "performance issues," "scalability issues", and "model degradation". As the AD area develops this shared language, the implicit discussions will become explicit. Another explanation for the high number of implicit studies is that AD issues are studied in isolation (e.g., hyperparameter tuning issues, feature engineering challenges, or model degradation) without recognising that these represent manifestations of the same underlying phenomenon. Thus, breaking down organisational and disciplinary silos is necessary for explicit AD recognition and management.

Building on this insight into the implicit framing of AD, we next consider how the underlying causes of AD reveal its multifaceted nature beyond surface-level terminology. One might expect AD to stem from a single root cause such as bad algorithm design, instead, we found three separate categories of data, algorithm design, and lack of ML knowledge. This highlights and positions AD as a multi-facet TD type consisting of both human and technical factors in developing ML/DL systems. These categories represent different levels of the ML/DL development stack. E.g., a developer may have access to clean data and understand algorithms well, but choose a suboptimal hyperparameter range due to time pressure. Hence AD emerges from the interaction of these interrelated factors, and not just any single cause.

Given the complexity of these interacting causes, examining how to address developer expertise is important. Although ML knowledge emerged as an important category of AD smells, surprisingly few studies explicitly discuss developer expertise gaps. This low number reflects a publication bias in ML/DL research, where technical innovations are emphasised while human and organisational constraints receive less attention. The limited availability of expertise-related issues suggests that current literature may understate a key driver of AD, particularly in industrial settings where practitioner skill levels vary widely. To address this gap, future work should investigate developer expertise directly through interviews, observational studies, or ethnographic methods, rather than relying solely on published research, which may systematically overlook human-centred contributors to AD.

### 5.6 Conceptual Framework

While we identified nine AD smells from this survey and their associated causes and consequences, the included studies did not explicitly report mitigation strategies. To address this gap, we propose a conceptual framework that extends the evidence by mapping potential remedies to the identified causes and consequences. This framework is a conceptual contribution, not a direct review result. It highlights how AD emerges from root issues in data, algorithm design, and ML knowledge, and how these propagate into inefficiency, scalability challenges, and model degradation (Figure 11).

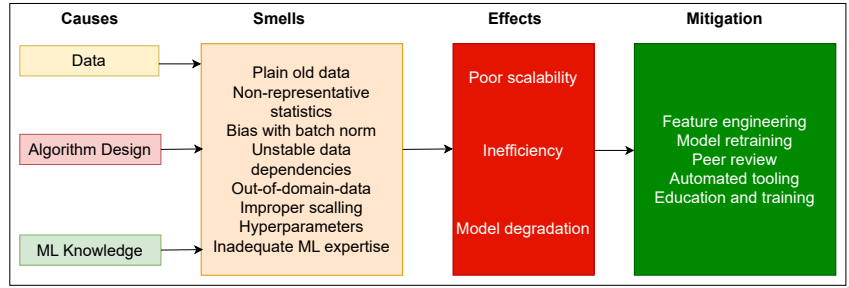

Fig. 11. Conceptual framework linking causes, smells, effects, and the proposed mitigation strategies for AD.

Having introduced the components and purpose of the conceptual framework, we now outline its AD mitigation strategies as follows: feature engineering for ML, model retraining, peer review, automated tooling, and education and training of practitioners. These strategies are derived from general ML/DL best practices and serve to illustrate possible pathways for addressing AD. By situating them alongside the smells and their consequences, the framework provides a practical guide for practitioners to anticipate and manage AD in ML/DL systems in relation to stated system goals.

## 5.7 Limitations

Our survey provides a comprehensive analysis of AD in ML/DL systems, but we note several limitations to guide future research. First, the scope of our survey may not fully capture AD in emerging ML/DL domains. We analysed 42 primary studies (Section 3), but this sample mostly represents academic perspectives, potentially missing AD manifestations in practice where practitioners rarely publish. E.g., RQ1 identified limited examples of scalability issues, potentially under-representing its impact. Future work should conduct practitioner-focused studies (e.g., industry case studies, field observations, and participatory workshops) to capture real-world AD, especially in large-scale production systems where scalability and degradation challenges are more visible with respect to AD.

Beyond the limitations of sample coverage, our analysis also highlighted interpretive constraints arising from the emerging nature of AD as a concept. Given that AD is an emerging concept rarely named explicitly in the literature, our implied interpretations of the original authors' texts may not always accurately reflect AD. This introduces a risk of misclassification. E.g., in our analysis of the primary studies, we inferred AD being discussed (i.e., training inefficiencies due to algorithm choices) in 83% (35/42 papers). This kind of interpretation might over or under attribute AD to phenomena like model training, potentially skewing our expanded definition (RQ1). Future validation studies could directly engage with the authors of the 35 inferred studies, e.g., through semi-structured interviews or follow-up surveys. Where author engagement is not feasible, we suggest that topic modelling or other natural language processing methods could be applied to systematically re-analyse textual evidence of AD-related constructs.

A further limitation concerns the variability in how the included studies conceptualised ML/DL systems. Although our inclusion criteria required some degree of system-level context, the reviewed papers still spanned different abstraction levels, from isolated ML modules embedded within larger workflows to full production pipelines. Such heterogeneity may influence how AD manifests or is prioritised in each context, as algorithm-level inefficiencies may be mitigated or amplified when integrated into broader systems. Future work might address this gap by developing a standardised taxonomy of ML/DL system boundaries and analysing AD manifestations across these tiers (e.g., algorithm- vs. pipeline- vs. deployment-level). A structured framework of this kind would enable more precise comparisons across studies and reveal whether certain AD smells are unique to specific system layers or propagate differently across the stack.

Finally, our focus on ML/DL systems may limit the applicability of our findings to other domains where algorithmic complexity is critical (e.g., game engines [49]). While our recommendations are tailored to ML/DL, our dataset was from DL frameworks, and their effectiveness in other contexts remains untested. This domain and context specificity may constrain the broader impact of our expanded definition of AD (RQ1). Future research should explore AD in diverse settings to verify the generalisability of our findings. This could also involve analysis on an expanded dataset that includes areas other than game engines where ML/DL is applied and beyond DL frameworks alone.

## 5.8 Threats to Validity

While the limitations reflect conceptual and scope-related constraints of the study, this section discusses methodological threats to validity related to study selection, search strategy, and data extraction. Validity threats arise from factors that may affect the accuracy and reliability of the findings [71].

To ensure the robustness of our search process, we addressed the bias in venue selection and took measures to minimise risks arising from the search string itself. As opposed to relying on a limited set of venues, we consulted multiple venues recommended in the academic literature [42]; ACM, IEEE, ScienceDirect, and Springer. We also used Google Scholar to complement the papers we obtained from these venues, by a snowballing process [41] to broaden coverage. When searching, we used the definition of ML/DL software from an existing study to find a scope for our work. We then identified terms, synonyms, and abbreviations from our control papers and validated the string by retrieving the studies on Google Scholar [61].

We also considered potential biases in data extraction and analysis. To mitigate the risk introduced by the use of one researcher during the extraction and analysis of data, the results were discussed and validated by an expert reviewer at each stage of the research process (e.g.,selection of the included studies and thematic analysis), ensuring a comprehensive and accurate representation of the extracted data and analysis [41]. Furthermore, coding of the AD categories performed by the first author was subsequently verified by an expert reviewer, thereby enhancing reliability. However, we did not compute a formal inter-coder agreement statistic. Future studies should employ such quantitative agreement measures (e.g., Cohen's Kappa [24] or Fleiss' Kappa [29] for two or more than two coders, respectively) to further strengthen reproducibility of the coding process.

## 5.9 Implications for ML/DL Practitioners and Researchers

Our findings highlight not only the presence of AD in ML/DL systems but also its practical and research implications. While prior sections have outlined the nature and manifestations of AD, here we translate these insights into actionable implications. Specifically, we provide guidance for ML/DL practitioners, who face immediate challenges in system design and deployment, and ML/DL researchers, who address conceptual and methodological gaps.

We present the following three implications for ML/DL practitioners: First, from a practical perspective, AD manifests as increased computational cost, reduced scalability, and repeated retraining cycles. The findings of RQ3 imply that suboptimal algorithm choices, unstable data dependencies, and other inefficiencies contribute to runtime and memory bottlenecks, particularly in large-scale and continuously deployed systems [18]. In production settings, such as autonomous or adaptive systems, these inefficiencies translate into AD as hidden operational costs, often measured in Graphics Processing Unit (GPU) hours and delayed deployment cycles [33]. This creates direct trade-offs between model performance, deployment frequency, and operational expenditure. Second, our findings suggest that MLOps pipelines' focus on model accuracy and deployment automation may be insufficient for managing AD. Practitioners should therefore view AD as a long-term maintainability concern rather than a short-term performance issue. Integrating AD-aware tests (e.g., automated validation of algorithmic efficiency, hyperparameter sanity checks, and detection of unstable data dependencies) into MLOps workflows should improve model maintainability by reducing the need for reactive refactoring and repeated retraining. Data (science) could mitigate AD arising from evolving or out-of-domain samples, which our findings identified as contributors to long-term degradation. Third, our results underscore the role of ML expertise in preventing AD. Limited familiarity with operations such as architecture selection, model optimisation, or efficient experimentation can introduce algorithmic inefficiencies that reduce ML/DL system maintainability over time. Investment in continuous upskilling and encouraging the use of best practices is therefore not only a human resource concern, but also a preventative strategy against long-term AD accumulation and maintenance overhead.

Our three implications for ML/DL researchers are as follows: First, the limited number of studies explicitly addressing AD (RQ2) indicates that AD remains under-theorised and frequently conflated with other TD types (e.g., Model, Data, or Design Debt). This presents an opportunity to more clearly delineate AD in ML/DL as a distinct research area based on algorithmic decision-making and ML-specific trade-offs. Second, our findings highlight opportunities for research on automated AD detection and its integration into MLOps workflows. We envision researchers leveraging advances in LLM-based SATD detection and pipeline instrumentation to develop automated AD detection tools embedded within continuous integration and deployment workflows. By operationalising AD in this way, the gap between conceptual understanding and practical ML/DL engineering could be bridged in the future, enabling proactive rather than reactive management of algorithmic inefficiencies. Third, our findings suggest that addressing AD may be particularly impactful in continual learning, real-time inference, adaptive systems, and other algorithm-intensive settings where model degradation unfolds gradually. Hence, longitudinal studies that track algorithmic decisions and their downstream effects over time would help clarify how AD accumulates and interacts with system evolution. By embedding AD considerations into both empirical research and tooling ecosystems, the research community can move toward more sustainable and maintainable ML/DL systems.

### 5.10 Research Roadmap

Based on the findings of RQ1–RQ3 and the limitations identified in this survey, we propose a phased research roadmap to advance the understanding and management of AD in ML/DL systems.

**Phase 1: Foundational Work.** The immediate priority is to formalise the boundaries between AD and related TD types such as Design, Model, and Data Debt. Findings from RQ2 shows that AD issues are misclassified as other TD types such as Design, Model, and Data Debt, while RQ1 and RQ3 highlight unique manifestations such as model degradation and inadequate ML expertise. To refine existing taxonomies, researchers should employ qualitative content analysis: systematically code TD instances from academic publications and open-source repositories using a rule-based codebook [53], and then engage multiple domain experts to classify each instance. Measuring inter-rater agreement will ensure reliability of distinctions between AD and TD types such as Model and Data that are conceptually similar. This systematic approach will yield a more precise understanding of the nuanced nature of AD in ML/DL systems to support targeted mitigation strategies tailored to its specific manifestations.

**Phase 2: Empirical Validation of AD Framework.** In addition to refining conceptual boundaries, future research should empirically validate our proposed framework (Figure 11) and the nine AD smells along with their prevalence and impact in real-world projects. We suggest a concise mixed-methods validation design. First, a structured online survey with ML/DL practitioners should be administered where they rate how often they encounter each of the nine AD smells and how strongly these relate to consequences such as degradation, inefficiency, and scalability issues, using Likert scales and open-ended prompts. This study should be followed with a smaller set of semi-structured interviews with about 15–20 ML/DL practitioners, to investigate how these smells actually arise in real projects and which mitigation strategies are applied, following established guidelines for industrial case studies in software engineering [92],

**Phase 3: Tool Development and Integration.** Complementing this validation effort, another promising direction would be to develop automated tools to detect and manage AD. To address ML knowledge gaps, future work should extend approaches from LLM-based SATD detection [23] to create targeted AD detection and remediation tools. Specifically, researchers should develop LLM agents that automatically identify AD patterns in code comments, commit messages, and configuration files by fine-tuning pre-trained models (e.g., transformer-based architectures) on an AD-specific corpus [77]. The detection pipeline should first classify whether a code segment contains an AD smell (e.g., hyperparameter misconfiguration or unstable data dependencies) using multi-class classification techniques, then generate context-aware remediation suggestions. To enhance the quality of recommendations, the system could further integrate knowledge of established hyperparameter optimisation strategies and architectural patterns for efficient ML/DL systems. Such tools would be particularly valuable for practitioners with limited ML expertise, addressing RQ3's ML Knowledge gaps, enabling proactive AD management during development and reducing accumulation of AD through informed decision-making rather than reactive fixes.

**Phase 4: AD Evolution and Management.** Extending all prior research directions, a longitudinal perspective is needed to understand how AD evolves over time in real-world ML/DL projects. How AD is introduced, accumulates, and is resolved over time in real projects remains unknown. We suggest that researchers conduct longitudinal studies on TensorFlow, PyTorch, or scikit-learn repositories to track when AD is introduced (e.g., which commits introduce inefficiencies) while also investigating how AD accumulates (i.e., does AD persist until major refactoring, or is it addressed incrementally?) and what factors influence its remediation (e.g., do community-driven projects resolve AD faster than corporate-maintained projects?) to guide practitioners resolve AD issues.

**Phase 5: Extension of AD Research.** Building on the validated AD framework, research should evaluate whether this framework generalises to complex systems beyond ML/DL, including, e.g., game engines, safety-critical control systems, and high-performance computing, by conducting domain-specific systematic reviews, repository mining, and expert validation. This would identify whether AD manifestations and root causes are generalisable or contextualised, e.g., by being domain-specific. Ultimately, this roadmap could transform AD from an underexplored TD type into an operationalised practice, enabling practitioners to identify algorithmic inefficiencies, organisations to implement proactive mitigation strategies, and tool developers to integrate AD detection into standard workflows.

## 6 Conclusions

In this study, we analysed 42 primary studies and an SATD dataset to advance the understanding of AD in ML/DL systems. We found three unique characteristics of AD in ML/DL systems which include model degradation, scalability limitations, and model inefficiency, which should be considered relative to system goals. Our findings show that AD is distinct, prevalent, and implicitly discussed in ML/DL systems. Our work makes several contributions. First, we provide a consolidated definition of AD in ML/DL systems characterising its unique manifestations. Second, our analysis quantifies the fragmented nature of current AD research, showing that only 17% of included studies explicitly mention AD, highlighting a gap in its awareness. Third, we identify nine smells that serve as practical indicators of AD. In addition, we propose a conceptual framework mapping the causes, effects, and potential mitigation strategies for AD, offering a structured guide for both researchers and practitioners.

These findings highlight that AD is not merely a technical deficiency, but a socio-technical phenomenon shaped by tooling, expertise, and organisational context. Addressing AD therefore requires not only improved algorithms, but also stronger software engineering practices, better developer education, and more supportive tooling ecosystems.

While we characterise what AD is, the empirical evidence of its evolution in practice and management strategies remain unknown. Moreover, our analysis, being largely based on academic literature, leaves industry-specific AD patterns and mitigation approaches largely unexplored. To advance the practical understanding of AD in ML/DL systems, we recommend that future research: i) validate our proposed AD framework and formalise the boundaries between AD and TD types such as Model, Data, and Design Debt, ii) develop automated detection tools leveraging recent advances in LLM-based SATD detection, and iii) conduct longitudinal case studies in production ML/DL systems to assess the real-world impact of AD and evaluate its mitigation strategies. Overall, this work establishes AD as a distinct and actionable research area and outlines a path toward its systematic study and mitigation in ML/DL systems.

## Acknowledgments

We want to thank Dr. Melina Vidoni for her valuable insights during the early stages of this work, which helped to shape the initial direction of our study. This work was supported by the Australian National University (ANU) through the ANU PhD scholarship within the ANU Research School of Computing, ANU College of Systems and Society.